\documentclass[10pt,letterpaper]{iopart}
\usepackage{graphics,color}

\newcommand{\ave}[1]{\left\langle #1 \right\rangle} %regular average
\newcommand \etc {{\it etc.} }
\newcommand \tie {{\it i.e.}}

\newcommand \kd  {\delta}
\newcommand \ra  {\rightarrow}

\newcommand \fn {{\bf n}}

\newcommand \si {\sigma}

\newcommand \p {^{\prime}}

\newcommand \lc {\langle}
\newcommand \rc {\rangle}
\newcommand \prt {\partial}

\newcommand \nt {\noindent}

\newcommand \bvec{\left( \begin{array}{c} }
\newcommand \evec{\end{array} \right)}

\newcommand \eg {{\it e.g.}}  
\newcommand \bea{\begin{eqnarray} }
\newcommand \eea{\end{eqnarray} }

\newcommand {\be} {\begin{equation}}
\newcommand {\ee} {\end{equation}}
\newcommand {\epem} {$e^+ e^-$}

\newcommand \mZ {{\mathcal Z}}
\newcommand \eqref[1] {(\ref{#1})}

\begin{document}

\article[Baryon number and strangeness]{Correlations and fluctuations 2005}
{Baryon number and strangeness: \\ signals of a deconfined antecedent} 
\author{A. Majumder, V. Koch and J. Randrup}
\address{Nuclear Science Division, 
Lawrence Berkeley National Laboratory\\
1 Cyclotron Road, MS:70R0319, Berkeley, CA 94720}
\begin{abstract} 
The correlation between baryon number and strangeness is used to 
discern the nature of the deconfined matter produced at vanishing 
chemical potential in high-energy nuclear collisions at the BNL RHIC. 
Comparisons of results of various phenomenological models with 
correlations extracted from lattice QCD calculations suggest that 
a quasi-particle picture applies.  At finite baryon densities, 
such as those encountered at the CERN SPS, it is demonstrated that the 
presence of a first-order phase transition and the accompanying 
development of spinodal decomposition would significantly enhance the 
number of strangeness carriers and the associated fluctuations.
\end{abstract}
%\preprint{LBNL-57969}

%\maketitle

%%%%%%%%%%%%%%%%%%%%%%%%%%%%%%
%%%%%%%%%%%%%%%%%%%%%%%%%%%%%%
%%%%%%%%%%%%%%%%%%%%%%%%%%%%%%

\section{Introduction}

%%%%%%%%%%%%%%%%%%%%%%%%%%%%%%
%%%%%%%%%%%%%%%%%%%%%%%%%%%%%%
%%%%%%%%%%%%%%%%%%%%%%%%%%%%%%

The goal of high-energy heavy-ion collisions is the creation of a new state 
of strongly interacting matter at very high energy density where the 
degrees of freedom carry color charges. These may be quark and 
gluon quasi-particles \cite{Collins:1974ky}, colored excitations 
of hadrons, or more complicated composite structures.   
The creation of such a state of matter is ushered in by the high 
energy densities created in such collisions. The reconversion 
of such an excited state of matter back into ordinary hadrons 
necessarily requires the presense of a phase transformation. The 
presence of such an excited state of matter and a phase change between 
it and ordinary matter 
has been observed in lattice simulations of QCD (LQCD) at a temperature of 170 MeV
and vanishing baryon density \cite{Karsch:2003jg}. 
The suppression of the number of high-momentum particles 
originating in the fragmentation of hard partonic jets requires 
the presence of a near opaque medium with densities at least 30 times 
that of conventional nuclear matter \cite{jet}. 
Such densities, when compared with the results from LQCD require the 
system to lie firmly in the deconfined state.

The phase transformation at 
a vanishing baryon density has been ascertained to be continuous. As a result, 
one would expect that the various conserved quantities such as the 
baryon number, charge and strangeness remain more or less unchanged 
within various sub-volumes of the phase space occupied by the system. 
As a result event-by-event fluctuations of these quantities may be used as 
probes of a state prior to hadronization. This will be the topic of 
the first part of these proceedings. It will be demonstrated that different 
assumptions regarding the nature of the Baryon number and strangeness carrying 
degrees of freedom lead to different measurable observables.

At zero temperature,
most models predict the occurrence of a first-order phase transition
when the density is raised. To date, there exist no 
lattice calculations in this regime.
Thus, if there exists a first-order phase transition
at zero temperature and finite chemical potential, 
this transition line should extend into the region of finite temperature
and would persist until the chemical potential has dropped
below a certain critical value $\mu_c$ \cite{Stephanov}.
There exist lattice QCD estimations \cite{fodor} that suggest the presence 
of a first-order phase transition line at finite chemical potential
as well as the existence of a critical end-point,
though its precise location is not well determined. The second part of 
these proceedings will deal with how certain observables such as the 
ratio of strange to non-strange particles and the fluctuations of 
such ratios are influenced by the non-monotonic behaviour of the 
thermodynamic potential as the system traverses a first-order phase 
transition line in the course of its dynamical expansion.

%%%%%%%%%%%%%%%%%%%%%%%%%%%%%%
%%%%%%%%%%%%%%%%%%%%%%%%%%%%%%
%%%%%%%%%%%%%%%%%%%%%%%%%%%%%%
%%%%%%%%%%%%%%%%%%%%%%%%%%%%%%

\section{Baryonic strangeness at RHIC: $\mu << T$}

%%%%%%%%%%%%%%%%%%%%%%%%%%%%%%
%%%%%%%%%%%%%%%%%%%%%%%%%%%%%%
%%%%%%%%%%%%%%%%%%%%%%%%%%%%%%
%%%%%%%%%%%%%%%%%%%%%%%%%%%%%%

Prior to the appearance of data from RHIC,
it was believed that results from LQCD could eventually be explained within a picture of 
the deconfined phase as a weakly interacting plasma of quasiparticles 
\cite{Andersen:1999va}.
The large radial and elliptic 
flow of the bulk matter however is somewhat inconsistent with the picture 
of a weakly interacting plasma \cite{Ackermann:2000tr}. 
Such observables seem to require the presense 
of a strongly interacting liquid. 
Indeed, recent results from lattice QCD calculation on spectral functions 
suggest the presence of bound, 
color-neutral states above $T_c$. 
This has led to the suggestion that at moderate temperatures, 
$T \simeq 1-2 \,T_c$,
the system is composed of medium-modified (massive) quarks and gluons
together with their (many) bound states \cite{Shuryak:2004tx}. 
In yet another picture, 
the system may be composed of strings between colored particles 
and their topological excitations \cite{Wang:1991ht}. 
At the present time, there exists no clear concensus 
on the nature of the degrees of freedom of the excited state of matter being created in 
RHIC collisions.

Imagine that highly excited matter has indeed been 
created in the mid rapidity region at RHIC. 
The matter is strongly interacting and as 
a result has thermalized in the course of 
its expansion. The thermodynamics of this 
system may, at this point, be described in terms 
of a temperature $T$ and 
chemical potentials $\mu_B, \mu_S, \mu_Q$ for each of 
the conserved quantities of baryon number, 
strangeness and electric charge \cite{Braun-Munzinger:2003zd}. 
Experimentally determined chemical freezeout 
temperatures place the chemical freeze out of the 
expanding gas of hadrons very close to the expected  
phase transition temperature. These features 
lead to the picture that the lifetime of 
the system as a gas of hadrons where the 
interactions are strong enough to change the chemistry 
of the state is rather short. It is this possibility 
which forms the basis of the assertion that 
the conserved quantities of baryon number $B$, 
strangeness $S$ and electric charge $Q$, 
within a wide rapidity bin will receive minor 
contamination from neighbouring bins.

As a result, the $B$,S and $Q$ established in 
a given rapidity bin in the plasma phase will be approximately 
maintained by the hadron gas phase in each 
event. 
It may be argued that 
the correlation between the strangeness $S$ and the baryon
number $B$ provides a useful diagnostic for the presence of 
strong correlations between quarks and anti-quarks.
Considering a situation in which the basic degrees of freedom
are non-interacting quarks and gluons,  the 
strangeness is carried exclusively by the $s$ and $\bar s$ quarks
which in turn carry baryon number in proportion to their strangeness,
$B_s=-\frac{1}{3}S_s$,
thus rendering strangeness and baryon number strongly correlated. 
This feature is in stark contrast to a hadron gas 
in which the relation between strangeness and baryon number is less intimate.
For example, at small chemical potential and temperature,
the strangeness is carried primarily by kaons which have no baryon number.

These considerations lead us to introduce 
the correlation coefficient \cite{Koch:2005vg},

\begin{equation}
C_{BS}\ \equiv\ -3 \frac{\ave{BS}-\ave{B}\ave{S}}{\ave{S^2} - \ave{S}^2}\
=\ -3 \frac{\ave{BS}}{\ave{S^2}}\ . \label{cbs_def}
%= - 3 \frac{\ave{ \delta B \delta S}}{\ave{(\delta S)^2}} ,
%-3 \frac{\ave{ (B - \ave{B})(S - \ave{S})}}{\ave{(S - \ave{S})^2}}
\end{equation}

\nt
The last expression uses the fact that $\ave{S}$ vanishes.
We have chosen the normalization such that $C_{BS}$ is unity
when the active degrees of freedom are individual quarks.

When the system consists of independent species $k$ 
having baryon number $B_k$ and strangeness $S_k$,
its total baryon number is $B=\sum_k n_k B_k$
and its total strangeness is $S=\sum_k n_k S_k$.
The correlation coefficient may then be expressed in terms of the multiplicity 
variances $\sigma_k^2\equiv\ave{n_k^2}-\ave{n_k}^2\approx\ave{n_k}$,

\begin{equation}
C_{BS}\ =\ -3 \frac{\sum_{k} \sigma_k^2 B_k S_k}{\sum_{k} \sigma_k^2 S_k^2}\
\approx\ -3 \frac{\sum_{k} \ave{n_k} B_k S_k}{\sum_{k} \ave{n_k} S_k^2}\ .
\end{equation}

\nt 
Thus, in the hadronic gas,
the numerator receives contributions
from only (strange) baryons (and anti-baryons),
while the denominator receives contributions also from (strange) mesons,

\begin{equation}
C_{BS}\ \approx\ 3{\ave{\Lambda} + \ave{\bar{\Lambda}} + \dots\,
+3 \ave{\Omega^-} + 3\ave{\bar{\Omega}^+} \over
K^0 + \bar{K}^0 + \dots\,
+9 \ave{\Omega^-} + 9\ave{\bar{\Omega}^+}} \, .
\end{equation} 

\begin{figure}[htb!]
 \hspace{-0.2cm}
  \resizebox{1.5in}{2.5in}{\includegraphics[0in,0in][6in,7.5in]{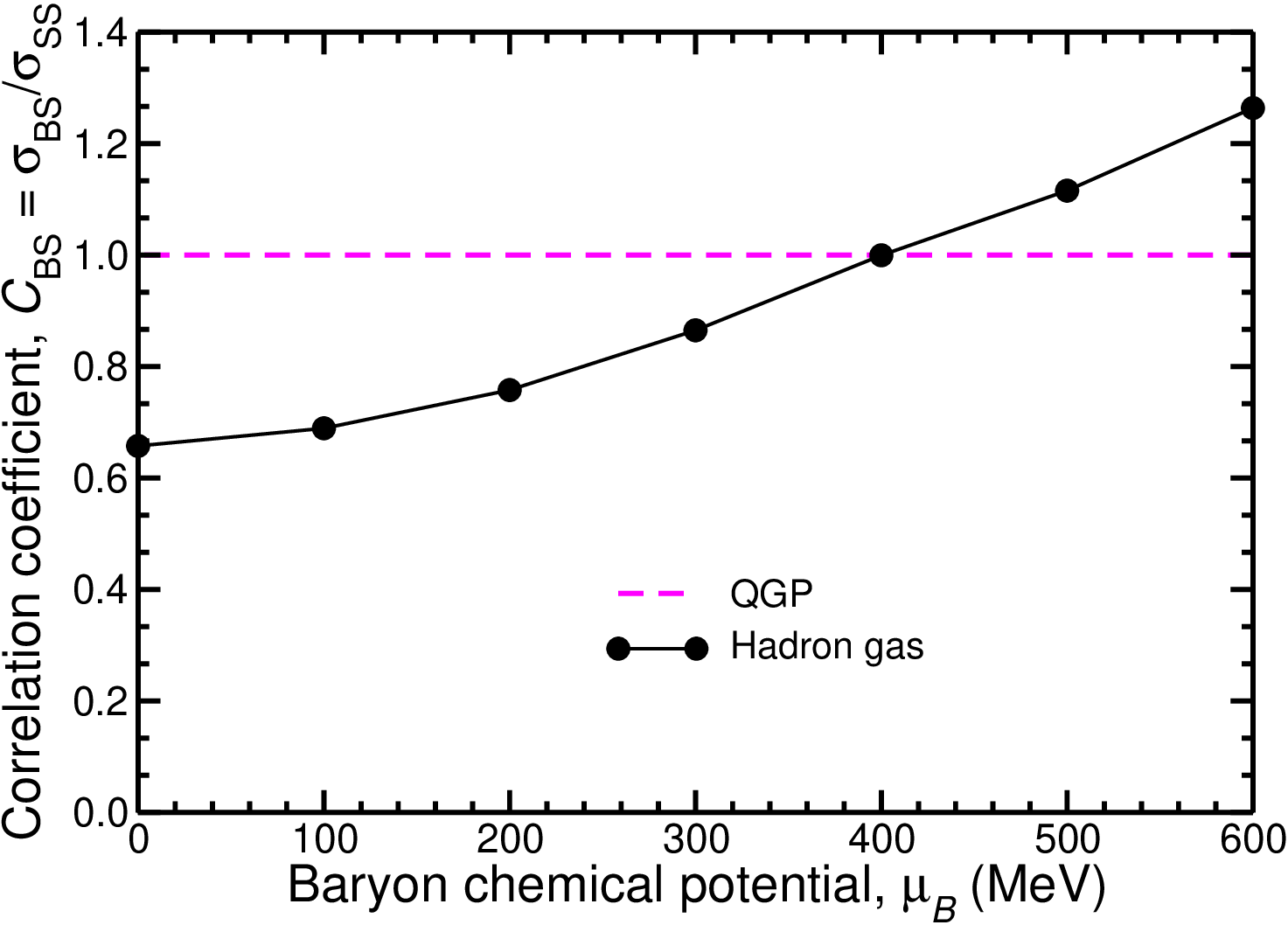}}
\hspace{2.8cm}
 \resizebox{1.5in}{2.5in}{\includegraphics[0in,0in][6in,7.5in]{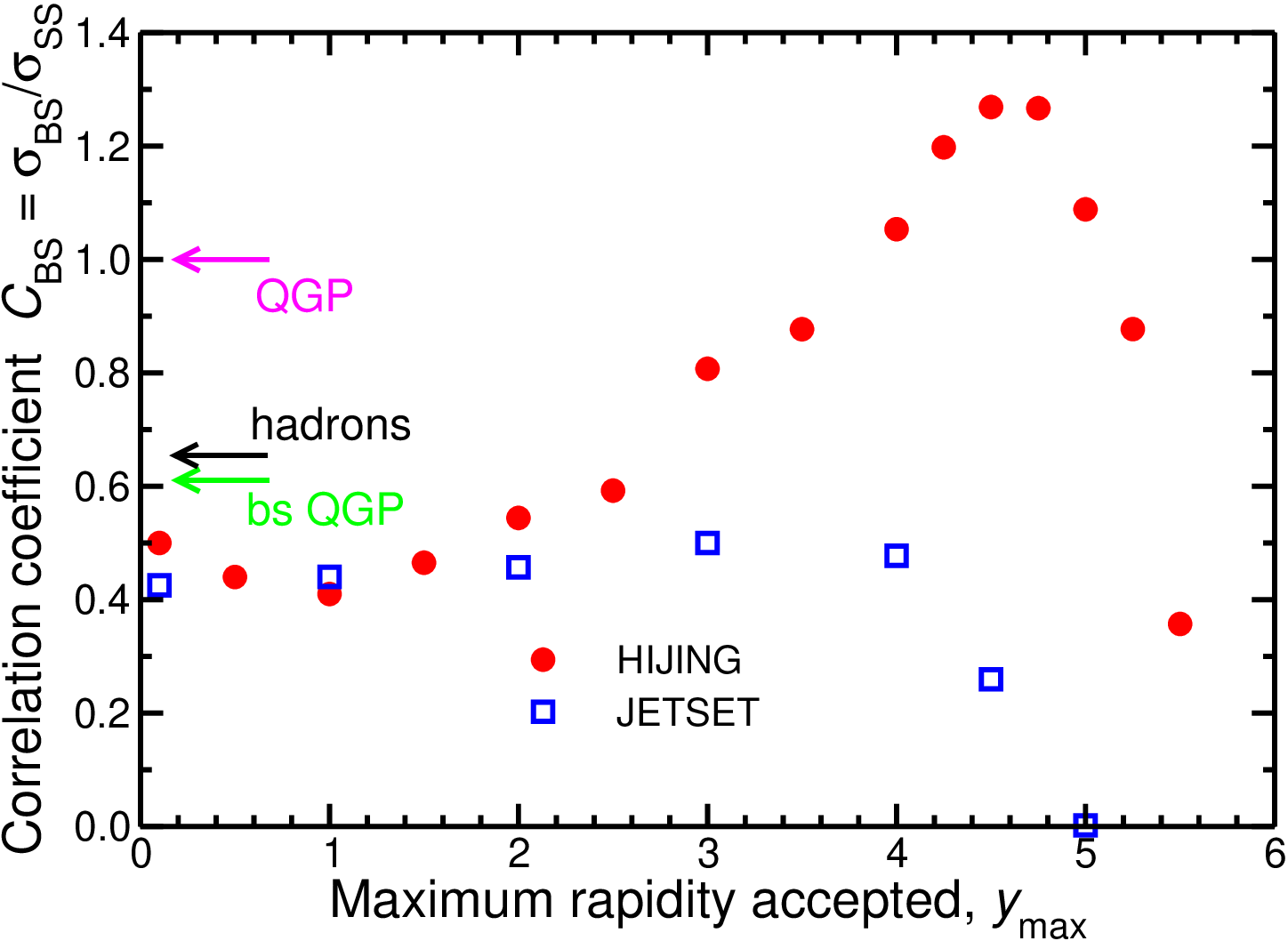}}
\caption{Calculations for the coefficient $C_{BS} = \lc \kd B \kd S \rc / \lc \kd S^2 \rc$ in 
various scenarios. The dot-dashed line represents $C_{BS}$ on the lattice and in a quasiparticle 
plasma; the dashed line is the estimate for a hadron gas and for a Shuryak-Zahed (SZ) plasma.
The filled circles represent the estimate from JETSET at center-of-mass energy = 20 GeV for 
various ranges of the maximum absolute rapidity of acceptance. 
The green triangles represent the results from HIJING 
simulations for Au-Au at 200AGeV for various ranges of acceptance. 
}
    \label{am-fig1}
\end{figure}

\nt
The hadronic freeze-out value value of $C_{BS}$ is shown in Fig.\ \ref{am-fig1}. 
At the relatively high temperatures relevant at RHIC,
the strange mesons significantly outnumber the strange baryons,
so $C_{BS}$ is smaller than unity.
Indeed, including hadrons up to $\Omega^-$,
we find $C_{BS}=0.66$ for $T=170~{\rm MeV}$ 
and zero chemical potential, $\mu_B=0$.

This is no longer the case when the baryon chemical potential is raised however. 
As an illustration, we trace the freeze-out line of Ref.~\cite{Becattini:2000jw}
which reports freeze-out temperatures and chemical potentials for a wide range 
of collision energies $\sqrt{s}$. As we go to lower energies the temperature decreases slightly 
however the chemical potential rises sharply; as a result the population of baryons 
rises in comparison to the mesons. This leads to an increase in the coefficient $C_{BS}$ as shown in the left panel of of Fig.~\ref{am-fig1} as 
a function of the chemical potential at the corresponding $\sqrt{s}$.

The value of unity for $C_{BS}$ in a quark-gluon plasma was 
obtained within 
the picture of a weakly coupled system. A more realistic value may be 
obtained from lattice simulations of QCD on the basis 
of off-diagonal susceptibilities \cite{Gavai:2002jt}. In these simulations, 
the partition function $Z$ for QCD with three quark flavors is estimated. The 
input parameters are the temperature $T$, and the three chemical 
potentials for the the up, down and strange flavors $(\mu_u,\mu_d,\mu_s)$.
Calculations are performed at vanishing chemical potentials. 

In this system the 
density of a flavor $f$ is obtained as the derivatives of 
$F = \log{Z} /V$ with respect to the chemical potential $\mu_f$.
The quark number susceptibilities are the second derivatives 

\bea 
\chi_{f f^{\p}} = T \frac{ \prt^2 F }{\prt \mu_f  \prt \mu_{f^{\p}}} \, .
\eea 

It may be easily argued 
that the mean flavour densities are zero at vanishing chemical 
potential for all flavours: $\lc u \rc = \lc d \rc = \lc s \rc = 0$. 
As a result, the coefficient $C_{BS}$ may be expressed in terms of the 
susceptibilities as

\bea
 -3 \frac{\lc \kd B \kd S \rc}{\lc \kd S^2 \rc} &=& 
 - \left. \frac{ \frac{\prt^2 Z(\mu_u,\mu_d,\mu_s)}
{ \prt \mu_B \prt \mu_s } }
{ \frac{\prt^2 Z(\mu_u,\mu_d,\mu_s)}{ \prt \mu_s^2 } } 
\right|_{\mu_u=\mu_d=\mu_s=0}
= 1 + \frac{\chi_{ds} + \chi_{us} }{ \chi_{ss} } \, .  
\eea
  
\nt
From Ref.~\cite{Gavai:2002jt} we obtain $\chi_{ss}/T^{2} = 0.53 (1) $ and 
$\chi_{us} + \chi_{ds} = 0.00 (3)$ at $T=1.5T_c$. As a result we obtain $C \simeq 1$ from 
lattice QCD, consistent with the estimate from a na\"{i}ve picture of a 
weakly interacting plasma. 
These results were obtained in a quenched approximation, 
but the effect of sea quarks is expected to be marginal above $T_c$
\cite{Gavai:2002jt}.
We may thus surmise that the lattice system has $C_{BS}\approx1$,
which is consistent with a plasma of quasi-particle quarks.

Recently there has appeared a model
that purports to explain both the equation of state 
as obtained on the lattice as well as the 
large flow observed in heavy-ion collisions 
\cite{Shuryak:2004tx}. 
The model describes the chromodynamic system
as a gas of massive quarks, antiquarks, and gluons together with
a myriad of their bound states generated by a screened Coulomb potential.
In order to assess the consistency of this model with lattice calculations, 
we estimate the coefficient $C_{BS}$ in such a scenario.

Our estimates are based on Ref.~\cite{Shuryak:2004tx} and the assumption that the 
fluctuation coefficient $C_{BS}$ is set at a temperature $T=1.5 T_c$. In this model, 
the plasma at a temperature 
$T_c < T < 3T_c$ possess massive quasi-particle excitations.  
The attraction between such colored states 
is modelled via a Coulomb potential with a Debye screening mass which is derived 
from lattice simulations. The temperature dependence of the quasiparticle and Debye 
masses is obtained by parameterizing data from the lattice. The attraction 
at such temperatures is still sufficient to produce a variety of bound states.
For three flavors of quarks there exist 749 such bound states. 
However, only the color-triplet $sg$ and the color-singlet $q\bar s$ states
(and their conjugates) are of relevance here.
(The color-hexaplet $sg$ states as well as the diquark states
are very weakly bound and dissolve entirely at $1.5\,T_c$.)
There are 2 $\pi$-like (spin-singlet)
and 6 $\rho$-like (spin-triplet) $q\bar s$ states, 
and 36 $sg$ states.  
The abundances of these states are estimated in a grand canonical ensemble
with vanishing chemical potentials.
The $q\bar s$ multiplets carry no baryon number 
and thus contribute only to $\sigma_S^2$,
hence drive $C_{BS}$ towards zero,
while more degenerate $sg$ multiplet contributes also 
to $\sigma_{BS}$ and thus drives $C_{BS}$ towards unity.
The resulting value obtained at $T=1.5T_c$, $C_{BS}=0.62$,
differs significantly from the value extracted 
directly from lattice QCD (see above),
thus suggesting that the bound-state model, does not 
encode the same degrees of freedom that are pervasive in lattice 
simulations of QCD.

Turning to more experimental considerations, we outline how such correlations 
may be actually measured in RHIC collisions. Such estimations are made with the 
aid of the Monte-Carlo event generator HIJING \cite{Wang:1991ht}.
The collisions occur in the center-of-mass frame at a $\sqrt{s}=200 AGeV$ 
and we estimate the 
coefficient $C_{BS}$ using the left hand side of Eq.~\eqref{cbs_def}. One 
measures all hadrons in a given event that lie within a range of rapidity 
from $-|y_{max}| < y_{accept} < |y_{max}|$, and computes the total baryon 
number $B$, the total strangeness $S$ and their products $BS$ and $S$. 
One then calculates the event averages of all these quantities and 
resurrects the coefficient $C_{BS}$. 
These results are represented by the closed circles in the right panel of Fig.~\ref{am-fig1}.

As the rapidity range of the acceptance is increased the coefficient remains more or less 
constant at a rather low value of 0.45. This is primarily due to the well know 
problem of baryon production in string fragmentation models. However as the
rapidity range is increased we note a large rise  in the coefficient. This is 
due to the fact that the net baryon number in the large rapidity regions is 
much larger than at mid-rapidity and as a result the number of baryons in the 
computation of the coefficient increases. This rise should be compared to the rise of the 
coefficient with baryon chemical potential in the left hand side of Fig.~\ref{am-fig1}. 
Eventually, at very large acceptances, one begins to capture the entire system. 
Due to the strict conservation of baryon number and charge in such models, the fluctuations 
of $BS$ drop rapidly as the acceptance encompasses the full system. There remains a
residual fluctuation in the strangeness content due to weak decays. As a result $C_{BS}\ra 0$
at large acceptance.

In order to study the effect of the exact conservation of baryon number without the 
influence of the rather large net baryon number, we repeat the above study on 
JETSET (\epem annihilation to two jets) events \cite{Andersson:1983ia} 
where the net baryon number is zero. 
As would be expected, 
at the lower rapidity range of the acceptance, it is consistent with the results from 
HIJING, which also displays a very small net baryon number at mid-rapidity. 
The coefficient $C_{BS}$ extracted from JETSET events 
shows a monotonous drop to zero as the acceptance is 
increased to encompass the full system. The value of $C_{BS}$ for asymptotically 
small ranges of the acceptance may estimated as simply the ratio of the probability to 
observe a strange baryon to that of strange meson.

In these proceedings, we have proposed an experimentally observed ratio 
$C_{BS} = -3 \lc \kd B \kd S \rc / \lc \kd S^2 \rc $, which may be used 
not merely to detect the prior existence of a chromodynamic phase of 
QCD, but indeed to decipher the strangeness and baryon number carrying 
degrees of freedom prevalent in such an environment. Estimates based on a 
quasi-particle picture of such degrees of freedom tend to favor a value 
of unity for the coefficient. This is also consistent with the estimates from lattice 
simulations. In a more strongly interacting scenario, which possesses light bound states 
of quarks and antiquarks the coefficient is reduced to approximately $2/3$ which is consistent 
with the coefficient from a hadronic resonance gas with no prior existence of a deconfined phase. 
Estimates  from Monte Carlo models of string fragmentation place the coefficient at an 
even lower value of 
0.45 nearly independent of the rapidity range of the acceptance up to $\pm2$ units of 
rapidity. Thus measurements of the 
coefficient $C_{BS}$ allow for a resolution of the baryonic and strange degrees of freedom in a 
quark gluon plasma. For further details see Ref.~\cite{Koch:2005vg}.

%%%%%%%%%%%%%%%%%%%%%%%%%%%%%%
%%%%%%%%%%%%%%%%%%%%%%%%%%%%%%
%%%%%%%%%%%%%%%%%%%%%%%%%%%%%%
%%%%%%%%%%%%%%%%%%%%%%%%%%%%%%

\section{Strangeness at the SPS: $\mu > T$}

%%%%%%%%%%%%%%%%%%%%%%%%%%%%%%
%%%%%%%%%%%%%%%%%%%%%%%%%%%%%%
%%%%%%%%%%%%%%%%%%%%%%%%%%%%%%
%%%%%%%%%%%%%%%%%%%%%%%%%%%%%%

In this section, the focus is turned to systems produced at the lower beam 
energies of the SPS. 
Deriving a freeze-out temperature $T$ and a chemical potential $\mu$ from 
fits to particle ratios leads to a $T \simeq 165 - 160 $ MeV and 
$\mu_B\simeq 240 - 300$ MeV for a collision energy $\sqrt{s} = 8-17 $ GeV. 
Most models \cite{Stephanov} indicate that at such large baryon 
chemical potentials the system may indeed undergo a first order phase transition 
from a partonic phase to a hadronic phase. As the collision energy is 
increased, the chemical potential tends to drop. As a result, the 
line of first-order phase transitions should eventually terminate in a 
second-order end point followed by the regime of continuous cross over as 
expected from lattice simulations at vanishing chemical potential. 
Current lattice estimates do indicate the presence 
of such a critical end point \cite{fodor}. 
The location of this point is however not 
well established. 

The proximity between the expected first-order phase transition 
line and the chemical freeze-out line at these collision energies 
suggests that a quark-gluon plasma could possibly be formed at these 
energies. This system will then traverse the phase transition line 
during its subsequent evolution into a hadronic gas. 
In this section, we outline the observational consequences of a first 
order phase transition 
on the multiplicity and fluctuations of strange particles
produced in such collisions.

A universal feature of first-order phase transitions
is the occurrence of spinodal decomposition,
which occurs as a result of the presence of a  a convex anomaly
in the associated thermodynamic potential.
This phenomenon occurs when bulk matter, by a sudden expansion or cooling,
is brought into the region of phase coexistence.
Since such a configuration is thermodynamically unfavorable
the uniform system prefers to reorganize itself into spatially separate
single-phase domains. These blobs of plasma immersed in an environment of 
hadrons tend
to have a characteristic scale \cite{Randrup:2003mu}.
The essential feature of such a scenario is that if the breakup
is sufficiently rapid, 
then whatever strangeness happens to reside within the region of the plasma
that forms a given blob will effectively become trapped and, consequently,
the resulting statistical hadronization of the blob 
will be subject to a corresponding canonical constraint on the strangeness.
This will lead to an enhancement of
the multiplicity of strangeness-carrying hadrons,
as compared to the conventional scenario where 
global chemical equilibrium is assumed to be maintained through freeze-out.
(This occurrence of an enhancement is qualitatively easy to understand,
since the presence of a finite strangeness in the hadronizing blob
enforces the production of a corresponding minimum number of strange hadrons.)
The {\em fluctuations} in the multiplicity of strange hadrons, such as kaons, 
are enhanced even more, thus offering a possible means
for the experimental exploration of the phenomenon.
Indications of an enhancement in the $K$-to-$\pi$ ratio 
has already been seen experimentally around $20-30~$ AGeV beam energy \cite{NA49_kp}. 
Furthermore, at the same energy, the fluctuation of this ratio
are strongly enhanced if expressed relative to mixed events.

Strangeness, as opposed to other conserved quantities such as charge and 
baryon number, is not brought in by the initial states in a heavy-ion 
collision. It is produced in the dense systems created in such collisions and 
thus the net strangeness in the full participant region is necessarily vanishing. 
There is however no constraint on the 
number of strangeness carriers \eg, the strange quarks and antiquarks or on the 
resulting  strange mesons and baryons.
For such an enhancement to occur the phase transition in question must necessarily 
be from a state with a greater number of strangeness carrying degrees of 
freedom (compared to the dominant carriers of entropy \eg, in a quark-gluon plasma) 
to a state with a fewer number of strangeness carriers as compared to the 
number of carriers of entropy (\eg, a hadron gas). 
The rapid expansion curtails the global chemical re-equilibration of the 
hadronic strangeness carriers emanating from different plasma blobs and results in 
a persistence of the overall enhancement of the mean number of strangeness carriers 
in the rarer hadronic phase.

As in most statistical estimations of the multiplicities of produced particles,
we consider a volume $V$ in which the produced strange quarks and 
antiquarks are assumed to reach full global chemical equilibrium (at plasma 
temperature $T_q$, and baryon and charge chemical potentials $\mu_B$, $\mu_Q$) 
under the canonical constraint of net total strangeness $S_{tot}=0$, \tie, we 
assume that the expansion is not rapid enough to offset the chemical or 
kinetic equilibrium of the produced quarks, antiquarks and gluons. The 
multiplicity distribution of the total number of strange quarks (identical to the 
distribution of anti-quarks) is given by the expression

\bea
P_{q}(N) =
P_{\bar{q}}(N) = \frac{\xi_q^N}{N!} \frac{\xi_{\bar{q}}^N }{N!} 
\left/ \sum_{N=0}^{\infty}  \frac{\xi_q^N}{N!} \frac{\xi_{\bar{q}}^N }{N!} \right. . \label{p_q_N}
\eea

\nt
The single quark (anti-quark) partition function $\xi$ is given as 

\bea
\xi_s &=& {g_s\over2\pi^2}{V T_q^3\over\hbar^3c^3}\tilde{K}_2({m_sc^2\over T_q})\,
 e^{(\mu_B B_s + \mu_Q Q_s)/T_q }\ .  \label{xi_s}
\eea

\nt
Where $g_s$ is the spin-color degeneracy of a strange quark and 
$\tilde{K}_2(x)= x^2 K_2(x)$ [$K_2(x)$ is the modified Bessel function of the second kind]. 
We denote the denominator of Eq.~(\ref{p_q_N}) which represents the canonical 
partition function of a system of strange quarks and antiquarks with net strangeness zero 
as $\mZ_0^{S_q=\pm 1} \equiv \mZ_0^q$.

Following the dynamical scenario outlined above, we imagine that the system 
decomposes into $p$ boxes of equal size $V_q = V/p$ 
(here and in the rest of this section we 
will use blobs and boxes interchangeably). The number of ways of distributing 
$N$ non-identical quarks into $p$ boxes with $n_1$ quarks in  box 1, $n_2$ in 
box 2 \etc, is given by the combinatorial factor,  

\bea
C_{\{ \fn \}} = \left. \frac{N!}{n_1! n_2! \ldots n_p!} 
\left( \frac{1}{p} \right)^N \right|_{\sum_i^p n_i = N} 
\eea 

\nt
In the above equation, $\{\fn\}$ refers to the set of occupation numbers of the 
various boxes or blobs \tie, $n_1,n_2,...n_p$.

With the aid of the above expression and Eq.~(\ref{p_q_N}) we obtain the probability 
distribution of events with quarks and anti-quarks distributed in the $p$ compartments 
according to the occupation vectors of quarks and antiquarks \tie, 
$\fn^q$ and $\bar{\fn}^{\bar{q}}$, as 

\bea 
P(\fn^q,\bar{\fn}^{\bar{q}}) &=&  \left. \prod_i^p \frac{\left( \frac{\xi}{p} \right)^{n_i}}{n_i!}
\left. \frac{\left( \frac{\xi}{p} \right)^{\bar{n}_i}}{\bar{n}_i!}
\right|_{\sum n_i = \sum \bar{n}_i} \!\!\!\!\!\!\right/  \mZ_0^q  \, . \label{p_n_nb}
\eea

These blobs of plasma now undergo expansion and hadronization. 
The final hadronic populations achieve chemical equilibration within 
a volume $ V_h \simeq \chi V_q$ (at a temperature $T_h$) 
in the vicinity of the now well separated plasma blobs. 
The expansion factor is estimated to be $\chi \sim 1.5 - 4$. 
As a result, the net strangeness of a blob $i$ ($s_i = \bar{n}_i - n_i$) is 
maintained post hadronization by the hadronic resonance gas. 
Thus a computation of the properties of the final hadronic gas 
requires the probability distribution of the net strangeness in the various blobs,

\bea
\frac{Q_p^0(s_1,...s_p)}{\mZ_0^q} = \!\!\!\!\!\! \sum_{n_1,...n_p,\bar{n}_i,...\bar{n}_p} 
\!\!\!\!\!\!P(\fn^q,\bar{\fn}^{\bar{q}})
\prod_i \kd(s_i + n_i - \bar{n}_i) \, .
\eea

\nt
In an experimental setup one never measures all the particles 
produced in the full phase space, but 
instead views a sub-sample of a given event. As a result, the 
focus will lie on the set of partially summed quantities,

\bea
Q_p^S(s_1,...s_k) &=& \prod_{i=1}^k Z^{\pm}_{s_i} 
\sum_{s_{k+1},...s_p} \prod_i Z^{\pm}_{s_i} \kd \left( \sum_{j=1}^p s_j - S \right)  \, .
\eea

\nt
Where $Z_{s_i}^\pm$ represents the canonical partition function of 
only strange quarks and antiquarks in a single blob $i$ with net 
strangeness $s_i$.
The above equation represents the probability distribution, that 
a system of strange quarks and antiquarks is distributed into 
$p$ separate compartments
and the strangeness in the first $k$ compartments is specified. 
The strangeness contents of the remaining compartments is summed 
over all allowed values.

Given the strangeness $S_0$ in a compartment, the distribution
of strange hadrons in that compartment may be computed using
the canonical partition function for strange hadrons including 
the six classes of such hadrons with strangeness $\pm1,\pm2,\pm3$ 
and with total 
net strangeness $S_0$,

\bea
Z^{\pm 1 \pm 2 \pm 3}_{S_0} = \prod_{S=\pm1,\pm2,\pm3} \sum_{N_S} 
{\zeta_S^{N_S}\over N_S!}
\delta \left( \sum_S N_SS-S_0 \right) \, . \label{canonical_part}
\eea

\nt
The effective single-particle partition function for the class 
of strange hadrons with strangeness 
$S$, $\zeta_S$, is a sum of the single-particle partition functions off 
all the hadrons belonging to this class. The general criteria for the inclusion 
a given strange hadron in the above equation will be that its mass be below $1680$ MeV and its 
decay width be below $200$ MeV.

The net strangeness $S_0$ in Eq.~\eqref{canonical_part} 
is given by the probability distribution
$Q_p^0(S_0)/\mZ_0^q$ computed above.
The chemical potentials involved in the
single-particle partition functions of the hadrons ($\zeta_k$)
are in principle also influenced by the value of $S_0$ in the given blob. 
One may also note that the non-strange sector will be influenced 
by the traversal of the system through the first order phase transition 
line. 
In these proceedings we will ignore such complications and refer the 
reader to Ref.~\cite{koch} for details. In what follows, we will 
assume the chemical potentials to be independent of the net strangeness 
in the blob. It will also be assumed that the non-strange hadron multiplicities and their 
fluctuations may be obtained by grand canonical estimates 
based on the measured freeze-out temperature and chemical potentials.

As a result of the above simplifications, we may obtain the mean and variance 
of particles of strangeness $k$ from a compartment of net strangeness $S_0$ as \cite{Majumder:2003gr}

\bea
 \mbox{\hspace{-1cm}}\lc n_{k} \rc^{S_0} = \zeta_{k}
\frac{Z_{S_0 - k}^{\pm 1, \pm 2, \pm 3}}{Z_{S_0}^{\pm 1, \pm 2 , \pm 3}}\, , \mbox{\hspace{0.4cm}}
(\si_k^{S_0})^2 &=& \zeta_{k}^2
\frac{Z_{S_0 - 2k}^{\pm 1, \pm 2, \pm 3}}{Z_{S_0}^{\pm 1, \pm 2, \pm 3}} 
+ \lc n_k \rc^{S_0} -(\lc n_k \rc^{S_0})^2.
\eea

\nt
The mean number of such particles from a single
blob produced in a scenario as outlined above may be obtained
by simply taking the convolution of the above equation with the probability
distribution of strangeness,

\bea
\lc n_{k} \rc  = \frac{\zeta_k}{\mZ^q_0} \sum_{S_0}
\frac{Z_{S_0 - k}^{\pm 1, \pm 2, \pm 3}}{Z_{S_0}^{\pm 1, \pm 2 , \pm 3}}
Q_p^0(S_0) \, .
\eea

\begin{figure}          %       -----------------------------------------
\hspace{-0cm}
\resizebox{2in}{1.9in}{\includegraphics[0in,1in][8in,7in]{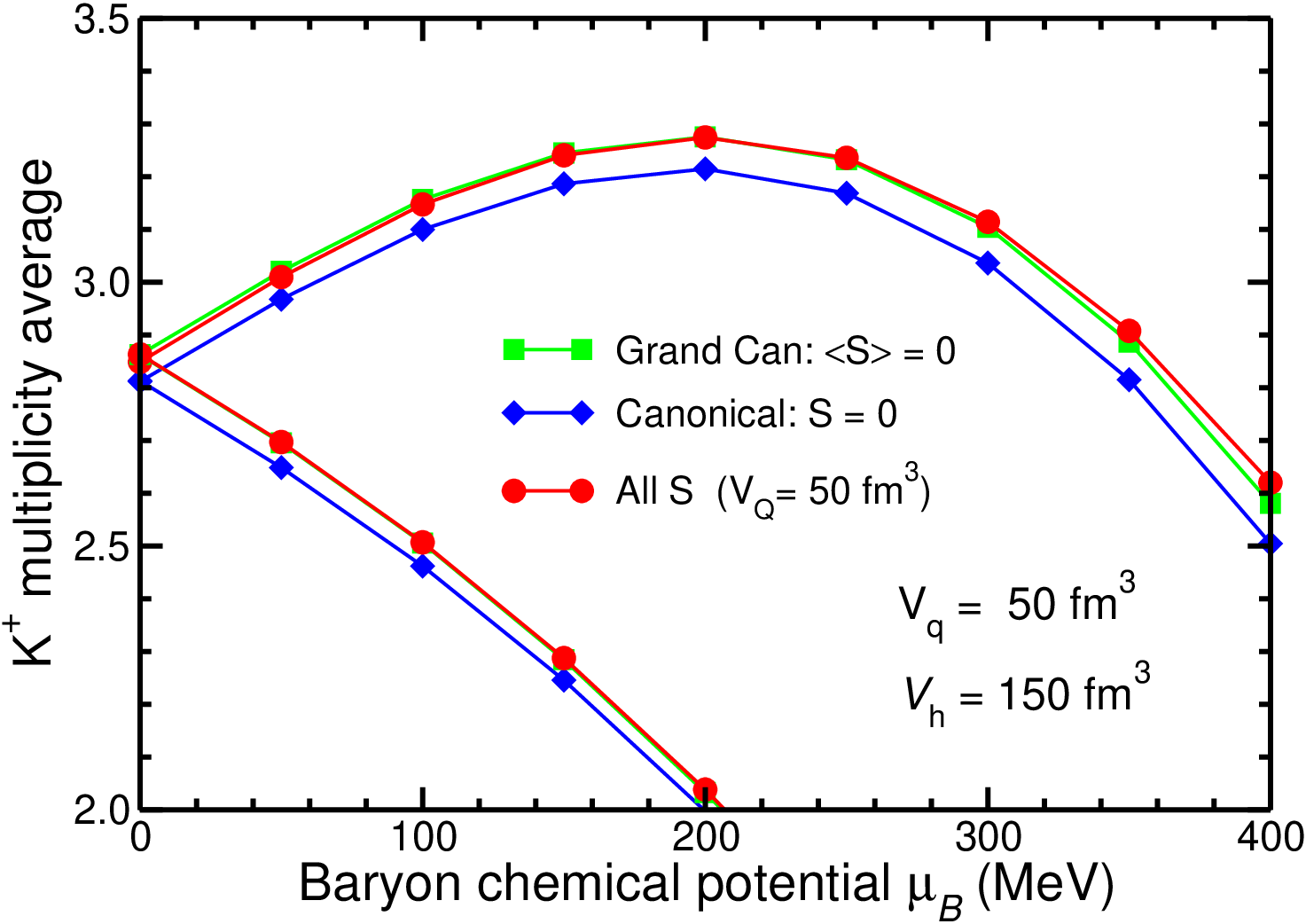}}	
\hspace{1.5cm}
\resizebox{2in}{1.9in}{\includegraphics[0in,1in][8in,7in]{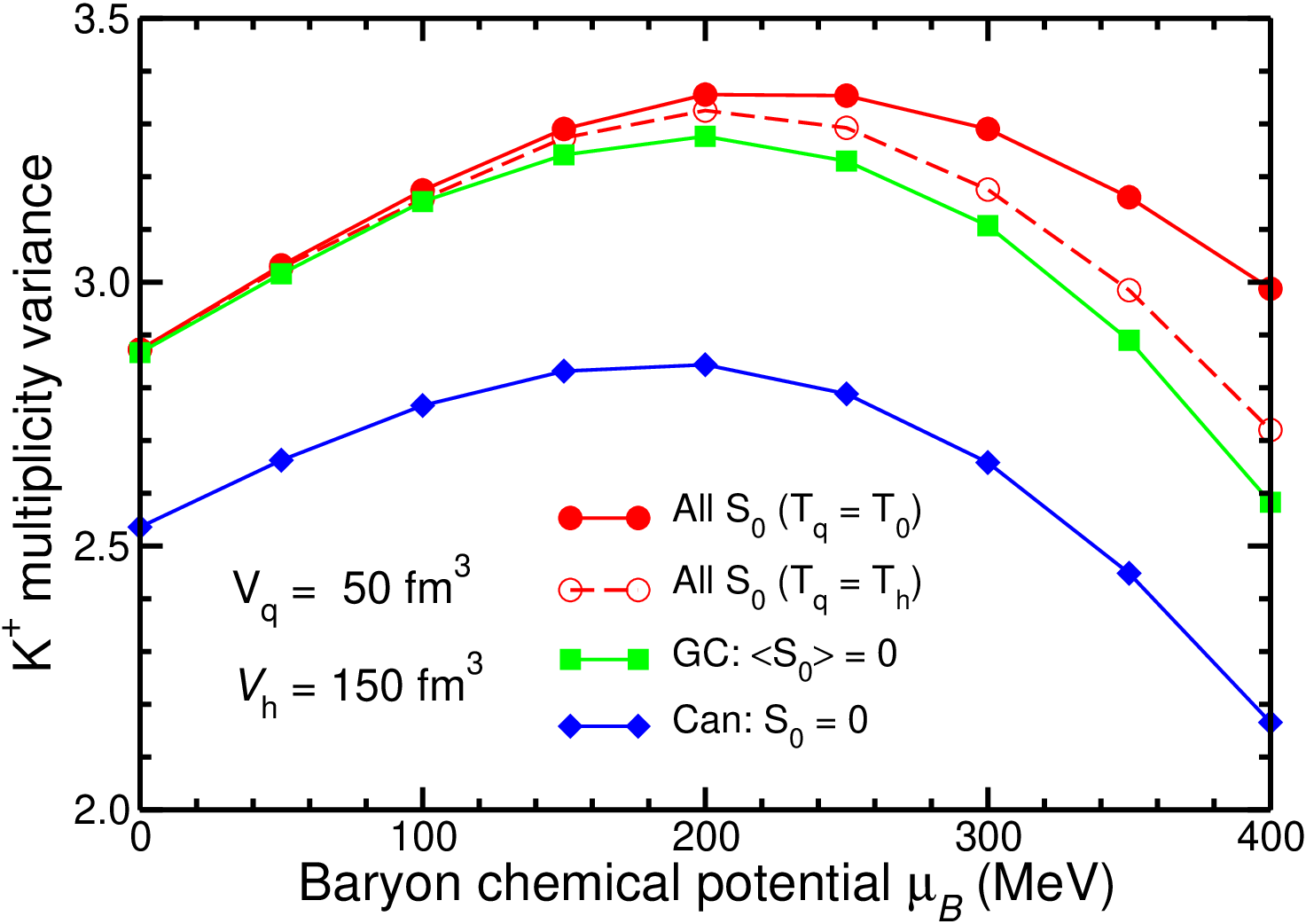}}
\caption{The average $K^\pm$ multiplicities and the $K^+$ variances 
as functions of the baryon chemical potential $\mu_B$
in three three scenarios:
1) the standard grand-canonical treatment,
2) the canonical treatment in which the blob strangeness $S_0$
is conserved through the hadronization
(the mean multiplicities are not sensitive to the plasma temperature $T_q$),
and 3) the restricted canonical treatment admitting only $S_0=0$.
}\label{am-fig2}
\end{figure}

As all the different blobs or compartments are considered equivalent,
we present results for the $K^\pm$ multiplicities and variances
from a single such blob of volume $V_q=50$ fm$^3$ in the plasma phase in Fig.~\ref{am-fig2}. 
For comparison, two other scenarios are 
presented. Besides the mean and variance that result from a spinodal 
decomposition during the phase transition, we also present the results 
from a simple grand canonical reference scenario.  
%The grand-canonical reference environment is determined as follows.
Using the baryon chemical potential $\mu_B$ as a control parameter,
we obtain the freeze-out temperature $T_h$ 
from the fit to the data obtained in Ref.\ \cite{Braun-Munzinger:2003zd}. 
Subsequently, we perform a grand-canonical iteration to determine 
those values of $\mu_Q$ and $\mu_S$ that ensure
$\bar{Q}=\alpha\bar{B}$ and $\bar{S}=0$,
where $\alpha=0.4$ which is representative of $Z/A$ for gold.
The last is referred to as the restricted canonical scenario where the 
blob strangeness is always fixed to zero.  

The general trend in all the three scenarios is similar. 
As $\mu_B$ is increased, the $K^+$ yield initially increases to balance the 
strangeness of the increasing number of strange baryons.
As the freeze-out temperature begins to decrease
noticeably, the hadron production generally decreases,
leading to a decreasing behavior of the $K^+$ curve. 
The multiplicities are insensitive to
the value adopted for the plasma temperature $T_q$,
which governs the fluctuations in the blob strangeness $S_0$. 
For the variances, the overall behavior is qualitatively similar
to the behavior of the averages, however, with some differences.
First, in the restricted scenario (where only $S_0=0$ is included)
the suppression of the variance is significantly larger.
This is the well know effect of canonical suppression. 
Furthermore, at the larger values of $\mu_B$,
where the net baryon density becomes significant,
the grand-canonical variance suffers more from the decreasing temperature
than the canonical variance.
This important divergence is a result of the fact that
the larger average baryon number implies a correspondingly larger
variance in the baryon number and therefore also a larger variance
in the strangeness. The enhancement is naturally reduced by reducing 
the plasma temperature as shown by the dashed line in Fig.~\ref{am-fig2}. 
In both cases, the results from the scenario with a 
first-order phase transition with strangeness trapping exceeds those 
from the other scenarios. This is especially true for the variance.

Work supported by the Director, Office of Science,
Office of High Energy and Nuclear Physics of the U.S. Department of Energy
(Contract DE-AC03-76SF00098).

\end{document}